 \documentclass[final,3p,times,twocolumn,number]{elsarticle}

\usepackage{amssymb}
\usepackage{lipsum}
 \usepackage{amsmath}
  \usepackage{xcolor}
\journal{Physics of the Dark Universe}

\begin{document}

\begin{frontmatter}

%% Title, authors and addresses

%% use the tnoteref command within \title for footnotes;
%% use the tnotetext command for theassociated footnote;
%% use the fnref command within \author or \affiliation for footnotes;
%% use the fntext command for theassociated footnote;
%% use the corref command within \author for corresponding author footnotes;
%% use the cortext command for theassociated footnote;
%% use the ead command for the email address,
%% and the form \ead[url] for the home page:
%% \title{Title\tnoteref{label1}}
%% \tnotetext[label1]{}
%% \author{Name\corref{cor1}\fnref{label2}}
%% \ead{email address}
%% \ead[url]{home page}
%% \fntext[label2]{}
%% \cortext[cor1]{}
%% \affiliation{organization={},
%%            addressline={}, 
%%            city={},
%%            postcode={}, 
%%            state={},
%%            country={}}
%% \fntext[label3]{}

\title{Notes on thermodynamics of Schwarzschild-like bumblebee black hole.}

%% use optional labels to link authors explicitly to addresses:
%% \author[label1,label2]{}
%% \affiliation[label1]{organization={},
%%             addressline={},
%%             city={},
%%             postcode={},
%%             state={},
%%             country={}}
%%
%% \affiliation[label2]{organization={},
%%             addressline={},
%%             city={},
%%             postcode={},
%%             state={},
%%             country={}}

\author[first]{Yu-Sen An}
\affiliation[first]{organization={College of Physics,\\Nanjing University of Aeronautics and Astronautics.},%Department and Organization 
            city={Nanjing},
            postcode={210016}, 
            country={China}}
\ead{anyusen@nuaa.edu.cn}
\begin{abstract}
%% Text of abstract
In this work, we revisited the thermodynamics of Schwarzschild-like bumblebee black hole using Iyer-Wald covariant phase space formalism. With a non zero vacuum expectation value, the bumblebee field is responsible for spontaneously breaking the Lorentz symmetry. As the bumblebee field is non-minimally coupled to gravity, we showed that the thermodynamic variables will be different from the counterpart in Einstein gravity. Especially, by using Iyer-Wald formalism, we found that the black hole entropy also differs from the result obtained from Wald entropy formula. Like Horndeski gravity, this mismatch is due to the divergence of bumblebee one-form field at the horizon. After figuring out the thermodynamics, we also briefly discussed the evaporation behavior of Schwarzschild like bumblebee black hole. We found that although bumblebee field has no influence on the critical impact factor, it can influence the black hole evaporation time. 
\end{abstract}

%%Graphical abstract
%\begin{graphicalabstract}
%\includegraphics{grabs}
%\end{graphicalabstract}

%%Research highlights
%\begin{highlights}
%\item Research highlight 1
%\item Research highlight 2
%\end{highlights}

\begin{keyword}
%% keywords here, in the form: keyword \sep keyword, up to a maximum of 6 keywords
Symmetry breaking \sep Black hole \sep Thermodynamics \sep Wald formalism

%% PACS codes here, in the form: \PACS code \sep code

%% MSC codes here, in the form: \MSC code \sep code
%% or \MSC[2008] code \sep code (2000 is the default)

\end{keyword}

\end{frontmatter}

%\tableofcontents

%% \linenumbers

%% main text

\section{Introduction}
\label{introduction}
General relativity and standard model of particle physics are two most successful theory which can provide description of the four fundamental interactions in nature. However,the pursuit of unification between the two is a very challenging task. Even though there are some candidate theory of quantum gravity, testing them in experiments or observations requires to reach the Planck scale ($\sim 10^{19}Gev$) which is far beyond current ability. However, there are some signals which may be observed at lower energy scales. An important one among them is the spontaneous breaking of Lorentz symmetry. 

The spontaneous breaking of Lorentz symmetry can be achieved by allowing the existence of a non-zero expectation value of some tensor field. One important model of them is the bumblebee model where a massive vector field couples non-minimally to the gravity sector \cite{Kostelecky:1988zi,Kostelecky:2003fs,Bluhm:2004ep,Kostelecky:2020hbb}. There are also constructions using Kalb-Ramond tensor field see Ref.\cite{Lessa:2019bgi} and reference therein. Previous studies constructed many solutions including stars,black holes and wormholes based on the bumblebee gravity theory \cite{Casana:2017jkc,Ovgun:2018xys,Maluf:2020kgf,Xu:2022frb,Filho:2024hri}. One simplest solution among them is the static spherically symmetric vacuum solution constructed in Ref.\cite{Casana:2017jkc}, which is called "Schwarzschild-like solution in bumblebee gravity" . This solution can be either seen as the exterior of a star or a black hole. In this paper, the author focuses on the advance of  perihelion and light bending aiming to shed light on its difference between Einstein gravity solution. The generalization to rotating case has also been done in Ref.\cite{Filho:2024hri,Ding:2019mal}. There are many further works exploring various aspects of this bumblebee gravity model, such as Ref.\cite{Lambiase:2023zeo,Li:2020dln,Mangut:2023oxa,Vagnozzi:2022moj,Uniyal:2022xnq,Kuang:2022xjp} for shadow and lensing properties of bumblebee black hole, Ref.\cite{Liang:2022gdk,Liang:2022hxd} for gravitational wave properties in bumblebee gravity, Ref.\cite{Gullu:2020qzu} for bumblebee black hole with topological defect and Ref.\cite{Khodadi:2022mzt} for cosmological implications of bumblebee vector field. 

Among these works, the thermodynamics associated to bumblebee black hole has not been fully understood. The previous works, such as \cite{Ding:2022qcy,Gomes:2018oyd} still use the same definition of thermodynamic variables as in Einstein gravity. While we should stress here that the thermodynamic variable depends heavily on the gravity theory. The standard definitions of Komar mass, ADM mass and the Bekenstein area formula for entropy only apply in Einstein gravity \cite{Wald:1999vt}. As there is non-minimal coupling between bumblebee field and gravity, the gravitational dynamics is different from Einstein gravity, thus the thermodynamic variables and the associated thermodynamic first law need to be reconsidered in bumblebee gravity model. 

Wald covariant phase space formalism \cite{Wald:1993nt,Iyer:1994ys} is the most general method to establish the thermodynamic laws of black hole. The advantage of Wald formalism is that it only relies on spacetime symmetries and is independent of the form of Lagrangian. So it is the more appropriate way for us to find the unambiguous form of thermodynamic variables for gravitational theories which are different from Einstein gravity. Thus there are many works discussing the black hole thermodynamics using Iyer-Wald formalism, to name a few see Ref.\cite{Bueno:2016ypa} for higher derivative gravity, and Ref.\cite{Feng:2015oea,Feng:2015wvb,Fan:2017bka}for matter field non-minimally coupled to gravity. Based on this reason, the Wald formalism will also help to elucidate the thermodynamic properties of bumblebee black hole. 

Throughout this work, in Sec.\ref{schbh}, we will first briefly introduce the Schwarzschild-like black hole in bumblebee gravity model. And in Sec.\ref{thermo},we will use Wald formalism to find the form of thermodynamic variables and establish the first law. After getting the thermodynamics of this black hole, we will investigate its evaporation behavior in Sec.\ref{evapo}. We end this work by concluding our result and discussing future directions in Sec.\ref{cd}.

\section{Schwarzschild-like black hole in bumblebee gravity}
\label{schbh}
In this section, we briefly review the solution obtained in Ref.\cite{Casana:2017jkc}.The bumblebee gravity model in 4d has the following action: 
\begin{equation}\label{action}
    S_{B}=\int d^{4}x \sqrt{-g}[\frac{1}{2\kappa}(R+\gamma B^{\mu}B^{\nu}R_{\mu\nu})-\frac{1}{4}B_{\mu\nu}B^{\mu\nu}-V(B^{\mu})]
\end{equation}
where $B_{\mu\nu}=\nabla_{\mu}B_{\nu}-\nabla_{\nu}B_{\mu}$, $\kappa=8\pi G$ and moreover we will set $G=1=c$ in the following. 

The specific expression of bumblebee potential is irrelevant \footnote{For example ,it can take the form $V(X)=\{\lambda X^{2}, \lambda X^{3}\}$ etc.} while in order to spontaneously break the Lorentz symmetry, there needs to be a non zero vacuum expectation value for the vector field, thus the potential should take the form 
\begin{equation}
    V=V(B^{\mu}B_{\mu}\pm b^{2})
\end{equation}
where $b^{2}$ is positive real constant. We consider the asymptotic flat spacetime which is free from cosmological constant, so the vacuum expectation value of bumblebee field is determined by 
\begin{equation}
    V(B^{\mu}B_{\mu}\pm b^{2})=0
\end{equation}
implying the condition 
\begin{equation}
    B^{\mu}B_{\mu} \pm b^{2}=0
\end{equation}
this is solved to be $B^{\mu}=b^{\mu}$ with $b^{\mu}b_{\mu}=\mp b^{2}$, $\pm$ sign determines whether $b^{\mu}$ is space-like or time-like. 

We are interested in the following metric ansatz and vector field ansatz \footnote{We only focus on the case where bumblebee field has only non-zero radial component, for the case where bumblebee field has non-zero time component see Ref.\cite{Xu:2022frb,Mai:2023ggs}}: 
\begin{equation}\label{ansatz}
\begin{split}
    &ds^{2}=-h(r) dt^{2}+\frac{dr^{2}}{f(r)}+r^{2}d\theta^{2}+r^{2}\sin^{2}\theta d\phi^{2},\\& B_{\mu}=b_{\mu}=(0,b(r),0,0).
\end{split}
\end{equation}
Here we consider spherical symmetric spacetime and also the vector field with only radial component obeying that 
\begin{equation}
    b_{\mu}b^{\mu}=b^{2}=const
\end{equation}
thus $b(r)=|b|/\sqrt{f(r)}$. Moreover, we are only interested in finding the simplest vacuum solution, where $V=V'=0$ and besides the bumblebee field ,there is no extra matter field. 

For this vector field ansatz with only radial component non-vanishing, we find 
\begin{equation}\label{condition}
b_{\mu\nu}=\partial_{\mu}b_{\nu}-\partial_{\nu}b_{\mu}=0.
\end{equation}
Varying the action Eq.(\ref{action}) with respect to the metric and bumblebee field gives the equation of motion.  By plugging in vacuum condition $V=V'=0$ and condition Eq.(\ref{condition}), the equation of motion gets simplified. Here we only quote the simplified equation of motion which reads 
\begin{equation}\label{einstein}
\begin{split}
    R_{\mu\nu}+&\gamma b_{\mu}b^{\alpha}R_{\alpha\nu}+\gamma b_{\nu}b^{\alpha}R_{\alpha\mu}-\frac{\gamma}{2}b^{\alpha}b^{\beta}R_{\alpha\beta}g_{\mu\nu}-\\&\frac{\gamma}{2}\nabla_{\alpha}\nabla_{\mu}(b^{\alpha}b_{\nu})-\frac{\gamma}{2}\nabla_{\alpha}\nabla_{\nu}(b^{\alpha}b_{\mu})+\frac{\gamma}{2}\nabla^{2}(b_{\mu}b_{\nu})=0
\end{split}
\end{equation}
\begin{equation}\label{beq}
    \nabla^{\mu}b_{\mu\nu}=-\frac{\gamma}{\kappa}b^{\mu}R_{\mu\nu}
\end{equation}
Plugging in the ansatz (\ref{ansatz}) leads to the following form of extended Einstein equation, 
\begin{equation}\label{eq2}
    (1+\frac{l}{2})R_{tt}+\frac{l}{2r}[\partial_{r}h(r) f(r)-\partial_{r}f(r)h(r)]=0
\end{equation}
\begin{equation}\label{eq1}
    (1+\frac{3l}{2})R_{rr}=0
\end{equation}
\begin{equation}\label{eq3}
    (1+l)R_{\theta\theta}-l[\frac{1}{2}r^{2}f(r)R_{rr}+1]=0
\end{equation}
where $l=\gamma b^{2}$.It can be easily found that Einstein equation Eq.(\ref{eq1}) implies $R_{rr}=0$. Thus from $b_{\mu}=(0,b(r),0,0)$, $b^{\mu}R_{\mu\nu}$ vanishes. Our ansatz of bumblebee field satisfies the bumblebee equation of motion (\ref{beq}). 

The equation of motion (\ref{eq2}),(\ref{eq1}),(\ref{eq3})can be solved analytically and a Schwarzschild like black hole can be found , which reads 
\begin{equation}\label{metric}
  ds^{2}=-(1-\frac{2M}{r})dt^{2}+\frac{l+1}{1-\frac{2M}{r}} dr^{2}+r^{2}d\Omega^{2}.
\end{equation}
The Kretschmann scalar can be easily calculated as 
\begin{equation}
R_{\mu\nu\rho\sigma}R^{\mu\nu\rho\sigma}=\frac{4(12M^{2}+4lMr+l^{2}r^{2})}{r^{6}(l+1)^{2}}
\end{equation}
which differs from Schwarzschild black hole and guarantees the black hole solution (\ref{metric}) is a true solution containing Lorentz violating correction. 
It is clear that the presence of $l$ breaks the Lorentz symmetry. The asymptotic spacetime structure fails to be the Minskowski spacetime because of this Lorentz violating factor $l$. 
In the next section, we will establish the thermodynamics of this black hole.

\section{Thermodynamics of Schwarzschild-like bumblebee black hole}
\label{thermo}
In this black hole spacetime, there exists a killing vector $\xi^{\mu}=(1,0,0,0)$ which is null at black hole horizon. As the theory is different from general relativity because of the non-minimal coupling between bumblebee vector field and metric field,  we must use Iyer-Wald formalism to establish the black hole thermodynamics. 

Firstly, either by using Euclidean method or quantum tunneling method \cite{Gomes:2018oyd}, the temperature of black hole is easily calculated as 
\begin{equation}
    T=\frac{\sqrt{f'(r_{h})h'(r_{h})}}{4\pi}=\frac{1}{4\pi r_{h}\sqrt{1+l}}
\end{equation} 
For black hole with the same horizon radius, we see the temperature of Schwarzschild-like bumblebee black hole is lower than the Schwarzschild black hole in Einstein gravity. 

For a Lagrangian in 4d spacetime,
\begin{equation}
    S_{B}=\int d^{4}x \sqrt{-g}L
\end{equation}
under the first order variation of dynamical field, the Lagrangian varies as
\begin{equation} \label{var}
    \delta L=E_{\phi}\delta\phi+d\Theta(\phi,\delta \phi)
\end{equation}
$E_{\phi}$ denotes the equation of motion, and $\Theta$ is the symplectic potential which is a three-form constructed locally from $\phi$ and $\delta \phi$. Consider the variation to be diffeomorphism $\delta \phi=\mathcal{L}_{\xi}\phi$, we find that 
\begin{equation}
    \mathcal{L}_{\xi}L=d(\xi \cdot L)
\end{equation}
where we use $dL=0$ because $L$ can be seen as a four-form on the four dimensional manifold.  From this the Eq.(\ref{var}) becomes 
\begin{equation}
    d(\xi \cdot L)=E_{\phi} \mathcal{L}_{\xi}\phi+d\Theta(\phi, \mathcal{L}_{\xi}\phi)
\end{equation}
It is easily seen that a Noether current can be defined by 
\begin{equation}\label{noe}
    J=\Theta(\phi,\mathcal{L}_{\xi}\phi)-\xi \cdot L
\end{equation}
thus $\mathrm{d}J=-E_{\phi} \mathcal{L}_{\xi}\phi$ which means that $J$ is a closed form when equation of motions are satisfied. As the closed form must be locally exact, there exists a two form $Q$, where 
\begin{equation}
    J=dQ
\end{equation}
A symplectic current can be constructed from $\Theta$ as
\begin{equation}
    \omega(\phi,\delta \phi, \mathcal{L}_{\xi}\phi)=\delta(\Theta(\phi,\mathcal{L}_{\xi}\phi))-\mathcal{L}_{\xi}(\Theta(\phi,\delta\phi))
\end{equation}
By doing the variation to Eq.(\ref{noe})
\begin{equation}
\begin{split}
    \delta J&=\delta[\Theta(\phi,\mathcal{L}_{\xi}\phi)]-\xi\cdot \delta L\\&=\delta[\Theta(\phi,\mathcal{L}_{\xi}\phi)]-\mathcal{L}_{\xi}[\Theta(\phi,\delta \phi)]+d(\xi \cdot \Theta(\phi,\delta \phi))
\end{split}
\end{equation}
So we obtain following relations: 
\begin{equation}
     \omega(\phi,\delta \phi, \mathcal{L}_{\xi}\phi)=\delta J-d(\xi \cdot \Theta)=d(\delta Q-i_{\xi}\Theta)
\end{equation}
By integrating above formula, we find 
\begin{equation}
    \int_{c}\omega(\phi,\delta \phi, \mathcal{L}_{\xi}\phi)=\int_{\Sigma}(\delta Q-i_{\xi}\Theta)=\delta H_{\infty}-\delta H_{+}
\end{equation}
where $c$ denotes Cauchy surface and $\Sigma$ is its boundary, which has two pieces, one is at asymptotic infinity and the other is at the horizon, and also $H$ is defined to be 
\begin{equation}
    \delta H_{\infty}=\int_{\infty}(\delta Q-i_{\xi}\Theta)
\end{equation}
\begin{equation}\label{deltahp}
    \delta H_{+}=\int_{r_{h}}(\delta Q-i_{\xi}\Theta)
\end{equation}
As $\xi$ is a killing vector, $L_{\xi}\phi=0$, $\omega(\phi,\delta \phi, \mathcal{L}_{\xi}\phi)$vanishes, which gives the relation 
\begin{equation}
    \delta H_{\infty}=\delta H_{+}
\end{equation}
The thermodynamic first law is the consequence of this formula.For spherical symmetric case without angular momentum,  $\delta H_{\infty}$ is identified as the variation of total spacetime energy which is defined at asymptotic infinity. It is called canonical Hamiltonian which will reduce to ADM Hamiltonian in Einstein gravity case. Meanwhile, $\delta H_{+}$ is identified as $T\delta S$ which is defined at black hole killing horizon, the entropy will have corrections compared to Bekenstein area law. 

In terms of our bumblebee gravity case (\ref{action}), Noether charge $Q$ \footnote{The variation with respect to bumblebee field $\mathcal{L}_{\xi}b^{\mu}$ vanishes} and $i_{\xi}\Theta$ reads
\begin{equation}
    \begin{split}
Q_{\alpha_{1}\alpha_{2}}=\epsilon_{\alpha_{1}\alpha_{2}\mu\nu} \{\frac{\partial L}{\partial R_{\mu\nu\rho\sigma}}\nabla_{\rho}\xi_{\sigma}&-\xi_{\sigma}\nabla_{\rho}(\frac{\partial L}{\partial R_{\mu\nu\rho\sigma}})\\&+\xi_{\rho}\nabla_{\sigma}(\frac{\partial L}{\partial R_{\mu\nu\rho\sigma}})\}
\end{split}
\end{equation}
\begin{equation}\label{itheta}
\begin{split}
   ( i_{\xi}\Theta)_{\alpha_{1}\alpha_{2}}=\epsilon_{\alpha_{1}\alpha_{2}\mu\lambda}\xi^{\lambda}\{2\frac{\partial{L}}{\partial R_{\rho\sigma\mu\nu}}\nabla_{\sigma}\delta g_{\rho\nu}&-2\nabla_{\nu}(\frac{\partial L}{\partial R_{\rho\mu\nu\sigma}})\delta g_{\rho\sigma}\\&+\frac{\partial L}{\partial(\nabla_{\mu}B_{\nu})}\delta B_{\nu}\}
\end{split}
\end{equation}
By using following identities,
\begin{equation}
    \frac{\partial R}{\partial R_{\mu\nu\rho\sigma}}=\frac{1}{2}(g^{\mu\rho}g^{\nu\sigma}-g^{\nu\rho}g^{\mu\sigma})
\end{equation}
\begin{equation}
\begin{split}
    \frac{\partial R_{\alpha \beta}}{\partial R_{\mu\nu\rho\sigma}}=X_{\alpha\beta}^{\mu\nu\rho\sigma}=&\frac{1}{8}\{(\delta^{\nu}_{\alpha}\delta^{\sigma}_{\beta}+\delta^{\sigma}_{\alpha} \delta^{\nu}_{\beta})g^{\mu\rho}-(\delta^{\mu}_{\alpha}\delta^{\sigma}_{\beta}+\delta^{\sigma}_{\alpha} \delta^{\mu}_{\beta})g^{\nu\rho}\\&-(\delta^{\nu}_{\alpha}\delta^{\rho}_{\beta}+\delta^{\rho}_{\alpha} \delta^{\nu}_{\beta})g^{\mu\sigma}+(\delta^{\mu}_{\alpha}\delta^{\rho}_{\beta}+\delta^{\rho}_{\alpha} \delta^{\mu}_{\beta})g^{\nu\sigma}\}
\end{split}
\end{equation}
and plugging in the action Eq.(\ref{action}) and field ansatz Eq.(\ref{ansatz}), we find\footnote{The last term in Eq.(\ref{itheta}) vanishes as $b^{\mu\nu}=0$} 
\begin{equation}\label{noet}
    Q=\frac{1}{16\pi}\{\sqrt{\frac{f}{h}}h'+l(\frac{1}{2}\sqrt{\frac{f}{h}}h'-2\frac{\sqrt{f h}}{r})\}
\end{equation}
    
\begin{equation}\label{itheta}
\begin{split}
    i_{\xi}\Theta&=\frac{1}{8\pi}(\sqrt{\frac{h}{f}}\frac{\delta f}{r}+\frac{1}{2}\sqrt{\frac{f}{h}}\delta h'+\frac{h'\delta f}{4\sqrt{fh}}-\frac{\sqrt{f}h'\delta h}{4h^{3/2}})\\&+\frac{l}{16\pi }(\sqrt{\frac{h}{f}}\frac{\delta f}{r}-\sqrt{\frac{f}{h}}\frac{\delta h}{r}+\frac{1}{2}\sqrt{\frac{f}{h}}\delta h'+\frac{h'\delta f}{4\sqrt{fh}}\\&-\frac{\sqrt{f}h'\delta h}{4h^{3/2}})
\end{split}
\end{equation}
By varying Eq.(\ref{noet}) and adding Eq.(\ref{itheta}) together, we find total contribution reads 
\begin{equation}
    \delta Q-i_{\xi}\Theta=-\frac{1}{8\pi r}\sqrt{\frac{h}{f}}(1+l)\delta f
\end{equation}
By plugging in Eq.(\ref{metric}), it is easily calculated 
\begin{equation}
    \delta H_{\infty}=\delta E=\sqrt{1+l}\delta M
\end{equation}
Also evaluating this on the horizon gives
\begin{equation}
    \delta H_{+}=T \delta S =\frac{\sqrt{1+l}}{2}\delta r_{h}
\end{equation}
It implies that we should define the thermodynamic variables of Schwarzschild-like bumblebee black hole as follows
\begin{equation}\label{entropy}
    E=\sqrt{1+l}M, \quad S=\pi r_{h}^{2}(1+l)
\end{equation}
and therefore the thermodynamic first law $dE=TdS$ holds. For $l \to 0$, it will reduce to the energy and entropy of Schwarzschild black hole. The energy, temperature and entropy satisfy following Smarr relation 
\begin{equation}
    E=2TS.
\end{equation}
The heat capacity of bumblebee black hole reads 
\begin{equation}
    C=\frac{\partial E}{\partial T}=-8\pi M^{2}(1+l)<0.
\end{equation}
The Lorentz violating factor will  affect heat capacity while the qualitative behavior of heat capacity in bumblebee black hole case is similar to Schwarzschild black hole which is thermally unstable. 

We conclude this section by discussing the form of energy and entropy (\ref{entropy}) we obtained. 

Firstly regarding the entropy, it is interesting to note that if we use Wald entropy formula \cite{Wald:1993nt} to compute black hole entropy which reads
\begin{equation}
    S_{W}=-2\pi \int d^{2}x\sqrt{-\gamma} \frac{\partial L}{\partial R_{\mu\nu\rho\sigma}}\epsilon_{\mu\nu}\epsilon_{\rho\sigma}.
\end{equation}
where $\epsilon_{\mu\nu}$ is the bi-normal associated to the horizon
\begin{equation}
    \epsilon_{\mu\nu}=k_{\mu}m_{\nu}-m_{\mu}k_{\nu}
\end{equation}
with $k_{\mu}=\{\sqrt{1-\frac{2M}{r}},0,0,0 \}$, $m_{\mu}=\{0,\frac{\sqrt{l+1}}{\sqrt{1-\frac{2M}{r}}},0,0\}$
Apart from the $A/4$ contribution ,there is an additional term due to the coupling between bumblebee field and Ricci curvature.  The Wald entropy formula gives the result
\begin{equation}
    S_{W}=\frac{A}{4}-\frac{\xi}{8} \int d^{2}x\sqrt{-h} X^{\mu\nu\rho\sigma}_{\alpha\beta}B^{\alpha}B^{\beta}\epsilon_{\mu\nu}\epsilon_{\rho\sigma}=\pi r_{h}^{2}(1+\frac{l}{2}),
\end{equation}
where we use the expression of integrand
\begin{equation}
X^{\mu\nu\rho\sigma}_{\alpha\beta}B^{\alpha}B^{\beta}\epsilon_{\mu\nu}\epsilon_{\rho\sigma}=-b^{2}. 
\end{equation}
It can be seen that the naive use of Wald entropy formula is different from the entropy we obtained in Eq.(\ref{entropy}) by carefully using Iyer-Wald formalism. 

The reason of this mismatch can be seen from the expression of Eq.(\ref{noet}) and Eq.(\ref{itheta}). From Eq.(\ref{noet}), it is easily seen that the Wald entropy formula is related to the Noether charge at the horizon, 
\begin{equation}
    \int_{r_{h}} Q \sqrt{-\gamma}d^{2}x= T S_{W}.
\end{equation}
Establishment of thermodynamic first law using Iyer-Wald formalism needs to do variation to the Noether charge at horizon, which is shown in Eq.(\ref{deltahp}) where $\delta H_{+}=T\delta S$ is identified to extract the expression of black hole entropy. In many gravity theories as was considered in Ref.\cite{Wald:1993nt,Iyer:1994ys}, the variation of Noether charge at horizon reads 
\begin{equation}
    \int_{r_{h}}\delta Q=T \delta S_{W}+S_{W}\delta T
\end{equation}
which can cancel with the terms in $\int_{r_{h}} i_{\xi}\Theta$ leaving precisely $T\delta S_{W}$ term. Thus $\delta H_{+}=\int_{r_{h}}(\delta Q-i_{\xi}\Theta)=T\delta S_{W}$ and therefore we find $S=S_{W}$. However, this coincidence no longer holds in the bumblebee black hole case. It can be seen that although the $\frac{-2l\sqrt{fh}}{r}$term in Eq.(\ref{noet}) vanishes at the horizon, its variation does not. Thus $\int_{r_{h}} (\delta Q-i_{\xi}\Theta)$ will have an additional contribution which is absent in $S_{W}$ rendering $S\neq S_{W}$ in bumblebee gravity model. The existence of the missing term relies on the fact that $l=\gamma b^{2}=\gamma g^{rr}b_{r}b_{r}\neq 0$, which can be attributed to the divergence behavior of bumblebee field $b_{r}=|b|\sqrt{\frac{1+l}{1-\frac{2M}{r}}}$ at the horizon. 

Note that this feature also appears in Horndeski black hole where the black hole entropy also does not coincide with Wald entropy formula \cite{Feng:2015oea,Feng:2015wvb}. The similarity between bumblebee gravity model and Horndeski gravity model can be seen from the Lagrangian. In Horndeski gravity, the Lagrangian is written as \cite{Feng:2015oea,Feng:2015wvb}
\begin{equation}
   L=\frac{1}{16\pi G}(R-2\Lambda)-\frac{1}{2}(\alpha g_{\mu\nu}-\gamma G_{\mu\nu})\partial^{\mu}\chi\partial^{\nu}\chi
\end{equation}
where scalar field $\chi$ has ansatz $\chi=\chi(r)$ and $g^{rr}\partial_{r}\chi \partial_{r}\chi \neq 0$ at horizon. Therefore, $\partial_{\mu}\chi$ behaves similarly as bumblebee field $b_{\mu}$ \footnote{$\alpha$ term in Horndeski gravity behaves like the potential of bumblebee field.}, and the bumblebee gravity model is similar to Horndeski gravity despite the fact that bumblebee field is non-minimally coupled to Ricci curvature tensor $R_{\mu\nu}$ instead of Einstein tensor $G_{\mu\nu}$.

As for the energy $E$, total energy of bumblebee black hole is no longer the same as mass parameter $M$. Furthermore, the gravitational action of bumblebee theory is different from Einstein gravity because of the non-minimal coupling between bumblebee field and metric, thus various energy definitions in Einstein gravity such as Komar mass may not work. As there exists time-like killing vector field, we can still define conserved current 
\begin{equation}\label{komar}
J_{R}^{\mu}=\xi_{\nu}R^{\mu\nu}
\end{equation}
and construct Komar-like mass from it. But this calculation will gives us 
\begin{equation}\label{komar}
    E_{K}=\frac{M}{\sqrt{1+l}} \neq E
\end{equation}
This mismatch between naive Komar mass Eq.(\ref{komar}) and canonical mass in Eq.(\ref{entropy}) indicates that the construction of conserved charge in Eq.(\ref{komar}) way fails to capture total spacetime energy.Thus we adopt the canonical mass which is derived by Wald formalism to be the spacetime total energy instead of Komar mass as adopted in Ref. \cite{Karmakar:2023mhs}. The readers can also consult \cite{Kastor:2008xb} for the construction of Komar-like mass in modified gravity theory. 

\section{Evaporation of Schwarzschild-like bumblebee black hole}
\label{evapo}
After getting the thermodynamics of the Schwarzschild like bumblebee black hole, we will also discuss other closely related thermodynamic phenomenon. Here we investigate one simplest phenomenon, the black hole's evaporation in geometric optics limit. There are many works investigating the evaporation of various kinds of black holes using this approach \cite{Page:2015rxa,Xu:2019wak,Liang:2023jrj,Wu:2021zyl,Hou:2020yni,Xu:2020xsl} 

From Stefan Boltzmann law, 
\begin{equation}\label{stb}
    \frac{dE}{dt}=T\frac{dS}{dt}=-a b_{c}^{2}T^{4}
\end{equation}
$a$ is some constant which will be set to 1 in the following discussion without loss of generality, and $b_{c}$ is the critical impact factor. Stefan-Boltzmann law will give the evolution of energy and horizon radius of bumblebee black hole. 

Firstly we should compute critical impact factor $b_{c}$ which is obtained from geodesic equation of motion. 

For a null geodesic , we can define two conserved quantity, the energy $E$ and angular momentum $L$ associated with two killing vector $K^{\mu}=(\frac{\partial}{\partial t})^{\mu}$, $\Psi^{\mu}=(\frac{\partial}{\partial \phi})^{\mu}$. 
\begin{equation}
E=-g_{\mu\nu}K^{\mu}U^{\nu}=(1-\frac{2M}{r}) \dot{t}
\end{equation}
\begin{equation}
L=g_{\mu\nu}\Psi^{\mu}U^{\nu}=r^{2}\dot{\phi}
\end{equation}
The trajectory of null geodesic (taking $\theta=\frac{\pi}{2}$) reads 
\begin{equation}
    (1+l)\dot{r}^{2}+(1-\frac{2M}{r}) \frac{L^{2}}{r^{2}}=E^{2}
\end{equation}
When considering the null particle radiates from horizon to infinity, we should demand the equation
\begin{equation}
(1+l)\dot{r}^{2}=E^{2}-L^{2}\frac{1-\frac{2M}{r}}{r^{2}}
\end{equation}
holds all along from horizon to asymptotic infinity. Define impact factor to be $b=L/E$, the above condition means for all radius position $r$,there is a relation
\begin{equation}
\frac{1}{b^{2}}\geqslant \frac{1-\frac{2M}{r}}{r^{2}}
\end{equation}
This gives the critical impact factor $b_{c}$, if $b<b_{c}$, the null geodesic can travel from horizon to infinity thus becomes a part of radiation observed from infinity. The critical compact factor is determined by the maximal value of effective potential $V(r)=\frac{1-\frac{2M}{r}}{r^{2}}$, 
\begin{equation}
    b_{c}^{2}=\frac{r_{p}^{2}}{1-\frac{2M}{r_{p}}}=27M^{2}=\frac{27}{4}r_{h}^{2}
\end{equation}
where $r_{p}=3M$ which is determined by $\partial_{r}V(r)=0$. 
It can be easily seen that the impact factor of bumblebee black hole is the same as Schwarzschild black hole in Einstein gravity. Thus from Stefan-Boltzmann law Eq.(\ref{stb}), the evolution of horizon radius obeys
\begin{equation}
    \frac{dr_{h}}{dt}=\frac{C}{r_{h}^{2}}
\end{equation}
where $C=-\frac{27}{512(1+l)^{5/2}\pi^{4} }$
By integrating the above formula we find that the total evaporating time reads 
\begin{equation}\label{evapt}
    t=\frac{512}{81}(1+l)^{5/2}\pi^{4}r_{0}^{3}
\end{equation}
where $r_{0}$ is the initial horizon radius. We see that in bumblebee black hole case, the relation between $t$ and $r_{h}$ remains similar, Lorentz violation parameter $l$ will only affect the black hole evaporation time by an overall factor. Compared with Schwarzschild black hole in Einstein gravity ,there is a relation
\begin{equation}
    t/t_{sch}=(1+l)^{5/2}.
\end{equation}
We also plot the evaporating behavior of bumblebee black hole for different Lorentz violating factor $l$ in Fig.1 
\begin{figure}[!ht]
	\centering	\includegraphics[width=0.55\textwidth,keepaspectratio]{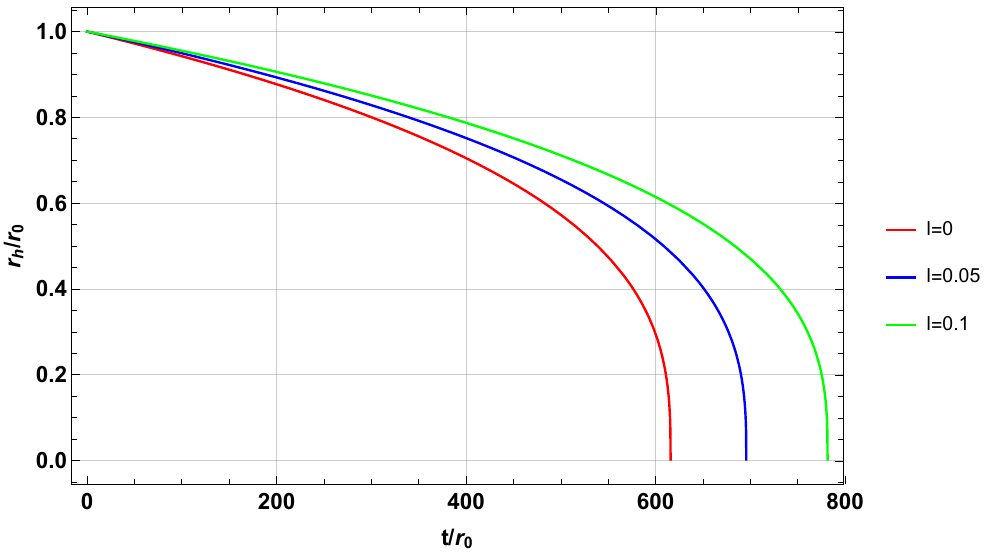}
 \caption{Relation between black hole radius $r_{h}$ and evaporating time $t$ during the evaporation process of bumblebee black hole.The red curve represents the Schwarzschild black hole case, and blue and green curve corresponds to $l=0.05$ and $l=0.1$ case respectively. Without loss of generality , we set the dimensionless constant $G/r_{0}^{2}=1$. We see the presence of Lorentz violating factor will enlarge the black hole evaporation time. } \label{3}
\end{figure}
It can be seen that the Lorentz violating factor can enlarge the black hole evaporation time.   The Lorentz violating factor is constrained to be very small using observations in the solar system such as $l\sim 10^{-12}$ by using perihelion data of Mars and Earth\cite{Casana:2017jkc,Pitjeva:2013xxa}.  But for the black hole far away from the solar system, there remains the possibility that $l$ will have relatively larger value and thus affect the black hole evaporation to a larger extent.

\section{Conclusion and Discussion}
\label{cd}
In this work, we have discussed the thermodynamics of Schwarzschild-like bumblebee black hole using Iyer-Wald formalism. Our result is different from previous works such as Ref.\cite{Ding:2022qcy,Gomes:2018oyd,Karmakar:2023mhs}. In these studies, the authors either identified mass parameter $M$ as spacetime energy or simply used Bekenstein-Hawking area formula $S=\frac{A}{4}$ as black hole entropy.\footnote{It was proposed there that the area should be redefined to $A \to \sqrt{1+l}4\pi r_{h}^{2}$.}  But as the action contains non-minimal coupling between bumblebee field and metric field, the gravitational dynamics is vastly different from the Einstein gravity. We can not expect the form of thermodynamic variables in Einstein gravity still hold in the bumblebee gravity model. We propose that the more appropriate way is to use Iyer-Wald covariant phase space formalism to find new expressions of energy and entropy and then establish the first law. We find that the expressions of internal energy and entropy are no longer the same as Einstein gravity counterparts. Interestingly, the entropy expression is also different from Wald entropy formula as a result of divergence of bumblebee field $b_{\mu}$ at the horizon. After getting thermodynamic first law, we also investigate other thermodynamic phenomenons such as black hole evaporation in geometric optics limit. 

There are many further directions which need to be discussed. 

Firstly, it is interesting to also compute the thermodynamic variables using Euclidean action and compare with the result obtained from Iyer-Wald formalism. For Euclidean method, the thermodynamic variable is related to Euclidean action by following relation 
\begin{equation}
    F=T I_{E}=E-TS.
\end{equation}
Thus the thermodynamic variables can be extracted from Euclidean action by using usual thermodynamic relation. Usually as shown in Ref.\cite{Iyer:1995kg}, the computations using Euclidean method and covariant phase space method coincide with each other. While there is a mismatch between the two in Horndeski gravity \cite{Feng:2015oea}.As bumblebee gravity model has many similarities with Horndeski gravity, there may also exist this mismatch in bumblebee gravity model. This mismatch may be related to the ambiguities in calculating Euclidean action where we should introduce counterterms in order to get finite Euclidean action as discussed in Ref.\cite{Feng:2015oea}. However, this question has not been fully understood yet. So it would be interesting to understand this phenomenon in a simpler model such as bumblebee model. This will offer further insight into the Euclidean approach to black hole thermodynamics.

Moreover, as there is construction of bumblebee black hole with cosmological constant \cite{Maluf:2020kgf}, it is interesting to go to extended phase space and investigate extended thermodynamics and associated phase transition behavior in bumblebee black hole. For extended phase space, the most important thing is to identify the expression of pressure. For asymptotically AdS spacetime, the most well-known proposal is to identify the cosmological constant as the pressure \cite{Kubiznak:2016qmn}. But as was shown in this paper, for bumblebee gravity model where the matter field is non-minimally coupled to gravity, using the expressions of thermodynamic variables of minimally coupled theory may cause problems. So it will be important to use more general method to find the expression of pressure such as using Iyer-Wald covariant phase space method. Recently, the extended Iyer-Wald formalism has been proposed to concretely find the expressions of thermodynamic variables and establish the extended thermodynamic first law \cite{Xiao:2023lap}. Thus it would be interesting to use this extended Iyer-Wald formalism to find the suitable definitions of pressure and volume in bumblebee gravity model. This will be helpful to further understand the phase transition behavior of bumblebee black hole.

\section*{Acknowledgements}
Yu-Sen An is supported by start-up funding No.90YAH23071 of NUAA. 

%% The Appendices part is started with the command \appendix;
%% appendix sections are then done as normal sections

%% If you have bibdatabase file and want bibtex to generate the
%% bibitems, please use
%%
\bibliographystyle{unsrt}
\bibliographystyle{elsarticle-harv} 
\bibliography{example}

%% else use the following coding to input the bibitems directly in the
%% TeX file.

%%\begin{thebibliography}{00}

%% \bibitem[Author(year)]{label}
%% For example:

%% \bibitem[Aladro et al.(2015)]{Aladro15} Aladro, R., Martín, S., Riquelme, D., et al. 2015, \aas, 579, A101

%%\end{thebibliography}

\end{document}